\documentclass[a4paper, 11pt]{article}
\usepackage{my_jcap} 

\usepackage{natbib}
\setcitestyle{aysep={}}
\usepackage{graphicx}	
\usepackage{amsmath}	
\usepackage{breqn}

\usepackage{fontawesome5}
\usepackage{color}
\usepackage{xcolor}

\allowdisplaybreaks  

\usepackage{mdframed}
\newmdenv[
    linecolor=gray,            
    backgroundcolor=gray!10,    
    linewidth=0pt,            
    roundcorner=0pt,            
    skipabove=\topsep,          
    skipbelow=\topsep,          
    innertopmargin=-5pt,         
    innerbottommargin=10pt,
    innerleftmargin=5pt,
    innerrightmargin=5pt,
    splitbottomskip=0pt,        
    splittopskip=30pt,           
    nobreak=false,               
]{mybox}

\newcommand{\msun}{\ensuremath{\, {\rm M}_\odot}} 
         
\newcommand{\mpc}{\ensuremath{\, {\rm Mpc}}}         
\newcommand{\gpc}{\ensuremath{\, {\rm Gpc}}}

\newcommand{\eg}{{\sl e.g.},\hskip 1pt}

\newcommand*{\vcenteredhbox}[1]{\begingroup
\setbox0=\hbox{#1}\parbox{\wd0}{\box0}\endgroup}

\newcommand{\OrcidID}[1]{ \href[urlcolor = red]{https://orcid.org/#1}{\textcolor{lightgray}{\faOrcid}}}
\newcommand{\OrcidIDName}[2]{\href{https://orcid.org/#1}{#2}}

\newcommand{\Rtwohc}{R_{\rm 200c}}
\newcommand{\Mtwohc}{M_{\rm 200c}}
\newcommand{\Vtwohc}{V_{\rm 200c}}
\newcommand{\Vmax}{V_{\rm max}}
\newcommand{\tVmax}{\tilde{V}_{\rm max}}

\newcommand{\fNL}{f_{\rm NL}}

\newcommand{\kOne}{k_1}
\newcommand{\kTwo}{k_2}
\newcommand{\kThree}{k_3}

\newcommand{\kvec}{\vec{k}}

\newcommand{\Bppp}{B_{\phi\phi\phi}}

\newcommand{\FFT}{\text{FFT}\,}
\newcommand{\iFFT}{\text{iFFT}\,}

\defcitealias{Scoccimarro2012PNGs}{S12}
\defcitealias{Sohn:2023:CMBbest}{S23}
\defcitealias{Sohn:2024:Colliders}{S24}
\defcitealias{paper1}{\textsc{Paper I}}
\defcitealias{paper2}{\textsc{Paper II}}
\defcitealias{paper3}{\textsc{Paper III}}

\usepackage[T1]{fontenc}   
\usepackage{lmodern}       
\usepackage{anyfontsize}   

\title{\fontsize{19.5pt}{24pt}\selectfont Primordial Physics in the Nonlinear Universe: \\Revealing the oscillating halo bias from cosmological collider models}

\author[1, 2]{\OrcidIDName{0000-0003-3312-909X}{Dhayaa Anbajagane}
(\vcenteredhbox{\includegraphics[height=1.2\fontcharht\font`\B]{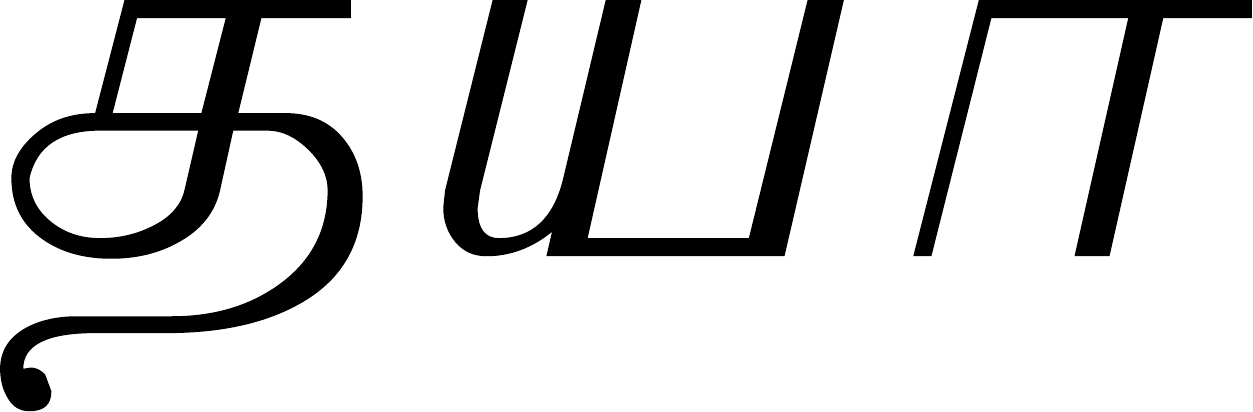}})}
\author[3]{and \OrcidIDName{0000-0001-9180-9726}{Neal Dalal}}

\affiliation[1]{Department of Astronomy and Astrophysics, University of Chicago, Chicago, IL 60637, USA}
\affiliation[2]{Kavli Institute for Cosmological Physics, University of Chicago, Chicago, IL 60637, USA}
\affiliation[3]{Perimeter Institute for Theoretical Physics
31 Caroline St. N, Waterloo, Ontario N2L 2Y5, Canada}

\emailAdd{dhayaacosmo@gmail.com}

\abstract{The initial conditions of our Universe contain a wealth of information about the particle physics of very high energies. One such class of signatures, called cosmological colliders, generates oscillations in the three-point correlations (or bispectra) of the primordial density field, and these imprint scale-dependent oscillations in the halo bias. We develop a new method for simulating cosmological collider models that foregoes traditional template-based basis-decomposition methods and can reproduce the scale-dependent signals of the input template at percent-level accuracy. Using this method, we produce simulations for one class of collider models and present the first measurements of oscillating halo bias in simulations. The amplitude and phase of the oscillations show a clear dependence on halo mass, with a factor of ten shift in halo mass causing a factor of two shift in the location of the oscillations. Increasing the frequency of the primordial bispectra model suppresses the signal in the halo bias, as the oscillations average down over the window function of the halo. The phase of the signal is also sensitive to assembly bias. In all cases, the scale-dependent halo bias can be accurately modeled using a simple peak background-split theory. The oscillations and their mass/selection-dependent phase offsets are a unique signature that is not easily mimicked by known observational systematics and is therefore a more robust target. Our simulations and underlying initial conditions code are both made publicly available.}

\makeatletter
\def\@fpheader{\ }
\makeatother

\begin{document}

\maketitle
\flushbottom



\section{Introduction}
Since the first high-resolution measurements of the Cosmic Microwave Background (CMB), as observed by the WMAP space mission \citep{Bennet:2003:WMAP}, the observational study of primordial non-Gaussianities \citep[PNGs,][]{Chen:2007:PNGs} has attracted considerable interest within the cosmology community. PNGs are features in the initial density field of our Universe that arise from non-trivial particle physics processes during the Universe's earliest epochs. As such, they provide a unique window into physics operating at very high energy scales \citep{Achucarro2022InflationReview}. Over the last decade, measurements from the \textit{Planck} space mission have significantly advanced our understanding of PNGs and placed tight constraints on the amplitudes --- commonly parameterized as $\fNL$ --- of many well-known PNG models \citep{Planck:2014:PNGs, Planck:2016:PNGs, Planck2020PNGs}.

While the space of PNG models can be vast, a large class of models has recently gained interest in the community. These models, referred to as cosmological colliders \citep{Nima:2015:Colliders}, arise from the interaction of the inflaton --- the particle governing the accelerated expansion during inflation; our leading paradigm for the early Universe \citep{Guth:1981:Inflation, Linde:1982:Inflation, Guth2004Inflation} --- with other particles that can be present at such high energies. These particles can have different masses, spins, and couplings with the inflaton, and as one varies the particles' properties, the PNG signatures can change \citep[\eg][]{Chen:2010:QSF, Baumann:2011nk, Sohn:2024:Colliders}. Thus, constraints on the amplitude of a PNG signature can directly inform the particle physics of the early universe. The key signature of cosmological collider models is oscillations in the three-point correlation of the initial density field. These correlations, written as the bispectrum $B(\kOne, \kTwo, \kThree)$ in Fourier space, exhibit oscillations in the squeezed limit of $\kThree \ll \kOne \approx \kTwo$ and generally scale as $B(k, k, q) \propto (k/q)^\Delta \cos(\log(k/q))$, where $q$ is the soft mode $q \ll k$. Since the bispectrum grows with $q \rightarrow 0$, these signatures peak on large scales. The observational program for cosmological collider physics is to measure or constrain the properties of these oscillations.

Thus far we have referred to the CMB as the key probe of inflation. However, the large-scale structure (LSS) of our universe, comprising cosmic structures like galaxies, galaxy clusters, voids, etc., can also constrain PNGs. In particular, the clustering of biased tracers (such as halos) was identified as a key probe of PNGs \citep{Dalal2008ScaleDependentBias}, because it exhibits unique scale-dependent features that cannot be mimicked by gravitational evolution. In particular, models of multi-field inflation predict a scale-dependent bias of $b \propto 1/k^2$ that grows towards larger scales. This signature is expected to enable constraints from galaxy surveys, which are probing increasingly large volumes of our Universe, to surpass the precision of those from the CMB \citep{Achucarro2022InflationReview}. Thus, the prospects are high for constraining these multi-field models of inflation. One would expect these prospects to also be shared by collider models, whose signals can similarly imprint on the halo bias and can often peak on large scales. However, the seminal work of \citet{Dalal2008ScaleDependentBias} has no direct, simulation-focused analog for the models of cosmological collider physics. More generally, there is no analog for PNG models with non-trivial behaviors in the squeezed limit. This is precisely the task we undertake in this paper.

A key challenge in performing such an analysis, and the reason it has not yet been pursued, is generating simulations with the correct bispectrum. The requisite method is now well-established for the widely studied class of inflationary bispectra \citep{Dalal2008ScaleDependentBias, Scoccimarro2012PNGs} but is not generalizable to other interesting bispectra, including the collider models we discuss here. A key issue is the requirement that the bispectrum be in separable form; we detail the precise requirements further below. \citet{paper1,paper2} provide a basis-decomposition approach that approximately decomposes the bispectra and produces simulations using this approximation, but their method is still limited by how well the chosen basis functions represent the input bispectrum. Here, we introduce a new method that does not rely on conventional basis functions for separability and is more flexible as a result. We use this new method to generate simulations of representative cosmological collider models, spanning multiple frequencies in their oscillations. Our focus is on measuring the non-monotonic, oscillatory features in the halo bias and fitting simple theoretical models to the measured behavior. Throughout, we detail the impact (or in some cases the lack thereof) of the unknown halo bias coefficients and assembly bias effects. 

This work is organized as follows: Section \ref{sec:sims} details our simulation and modeling methodology, including a discussion of our new initial conditions generator. We present our key results in Section \ref{sec:results}, and conclude in Section \ref{sec:conclusions}. Appendices \ref{appx:bias} and \ref{appx:local} present the best-fit bias coefficients, and a validation of our method for simpler PNG models, respectively. All analyses we show assume the fiducial cosmology used in the \textsc{Quijote} simulations \citep{Navarro2020Quijote}; $\Omega_{\rm m} = 0.3175$, $\sigma_8 = 0.834$, $n_s = 0.9624$, $h = 0.6711$, and $\Omega_{\rm b} = 0.049$.

\section{Simulations \& Modeling}\label{sec:sims}

Producing simulations of a given PNG model requires stringent constraints on the mathematical representation of a given PNG model (i.e. of the bispectrum), and this makes it challenging to generate initial conditions (ICs) for arbitrary PNGs. In \citet[][henceforth \citetalias{paper1}]{paper1} and \citet[][henceforth \citetalias{paper2}]{paper2}, we built a new basis decomposition method for tackling this issue. However, both works noted that the performance of the method was set by the appropriateness of the chosen basis. In this work, we are focused on accurately simulating the strongly squeezed limit of collider models, which is not always well represented by the basis decomposition methods. As a result, we introduce a new method for ICs that has superior reliability, particularly at the edges of the 3D space occupied by $\kOne, \kTwo, \kThree$. We first detail our methodology in Section \ref{sec:sims:ICs} and describe the model we consider in Section \ref{sec:sims:cosmocollider}. Section \ref{sec:sims:simconfig} lists the configuration of the simulations we run for this work. Finally, in Section \ref{sec:sims:theory} we detail the theory model used to fit the measurements from the simulations.

\subsection{Initial Conditions}\label{sec:sims:ICs}

The non-Gaussian inflaton potential, $\phi_{\rm NG}$, can be generated as
\begin{equation}\label{eqn:conv}
    \phi_{\rm NG}(\kvec) = \phi_{\rm G}(\kvec) + \fNL\int d^3k_1 d^3k_2\, (2\pi)^3 \delta(\kvec - \kvec_{12}) K_{12}(\kvec_1, \kvec_2) \phi_{\rm G}(\kvec_1)\phi_{\rm G}(\kvec_2) 
    \,,
\end{equation}
where $\phi_{\rm G}$ is the Gaussian inflaton potential, $\vec k_{12}  = \vec k_1 + \vec k_2$ and $ K_{12}(\kvec_1, \kvec_2)=K_{12}(\kOne, \kTwo, \kThree)$ is a mode-coupling kernel that is derived from a given bispectrum. We will discuss the form of the latter later in this section. Throughout this work, we assume the ordering $\kOne \geq \kTwo \geq \kThree$. The direct evaluation of the above integral has a poor computational complexity (scales as $N_{\rm grid}^6$) and is unfeasible for even modestly sized grids. \citet{Scoccimarro2012PNGs} showed this integral can be done efficiently using Fast Fourier Transforms (FFTs) if the kernel is separable in $\kOne, \kTwo, \kThree$. That is, if we split the kernel as $K_{12}(\kOne, \kTwo, \kThree) = f_1(\kOne) f_2(\kTwo) f_3(\kThree)$ then the above integral can be evaluated as
\begin{equation}\label{eqn:sim:ICs:S12}
    \phi_{\rm NG}(\vec{k}) = \phi_{\rm G}(\vec{k}) + \fNL\bigg[f_3(k_3) \times \FFT\Big\{\iFFT \Big(f_1(k_1) \phi_G(k_1)\Big) \times \iFFT \Big(f_2(k_2)\phi_G(k_2)\Big)\!\Big\}\!\bigg],
\end{equation}
following the approach of \citet{Scoccimarro2012PNGs}. Here, we have denoted the inverse FFT operation as $\iFFT = \FFT^{-1}$. The scaling of the above algorithm goes as the scaling of a 3-dimensional FFT, $N_{\rm grid}^3 \log(N_{\rm grid})$. This is a feasible calculation even for large grids. Thus, the operative challenge is taking the kernel, $K_{12}(\kOne, \kTwo, \kThree)$ and representing it in separable form.

In \citetalias{paper1} and \citetalias{paper2}, we utilized a basis decomposition method to approximate a given kernel with a separable form. While this method provided sufficient accuracy for a variety of models, the above works noted that the method was unable to reproduce a subset of models that exhibited more complicated behavior in their bispectra (e.g., high-frequency oscillations). The reliability of this approach depends on the choice of basis functions and how well those functions match the target template. \citet{Philcox:2025:ML} developed a Neural Network-based method for generating arbitrary basis functions to approximate a given target template. This method is precise for analyses of CMB observables, but we found it challenging to make the approximation sufficiently precise for large-scale structure, where specific subdomains of the 3D space --- e.g., the squeezed limit --- can dominate the late-time evolution while only making up a small fraction of the function space.

In this work, we forego the use of analytic (or Neural Network-based) basis functions. We instead utilize ``binning'' as a separable representation of the kernel. With the binned estimator approach, we can write the kernel as,
\begin{equation}\label{eqn:kernel:heaviside}
     K_{12}(\kOne, \kTwo, \kThree) = \sum_{ijl} \alpha_{ijl} \Theta_i(\kOne) \Theta_j(\kTwo) \Theta_l(\kThree)
\end{equation}
where the three Heaviside functions identify a given bin in 3D Fourier space, and $\alpha_{ijl}$ is the value of the kernel at the center of the bin. The above is the simplest iteration of a binned estimator as it has sharp edges. However, we can extend it to allow for smoother transitions across bins by using a higher-order kernel,
\begin{equation}\label{eqn:kernel:cubic}
K_{12}(\kOne,\kTwo,\kThree) = \sum_{i j l} \alpha_{ijl}\, \psi_i(\kOne) \psi_j(\kTwo) \psi_l(\kThree),
\end{equation}
where $\psi$ is a third-order (i.e., cubic) Lagrange interpolator \citep{NumericalRecipes} of the form,
\begin{equation}\label{eqn:psi_lagrange}
\psi_i(k) =
    \begin{cases}
    \displaystyle
    \prod_{\substack{m\in \mathcal S(k),\\ m\neq i}}
    \frac{k - \hat{k}_m}{\hat{k}_i-\hat{k}_m}, & i\in \mathcal S(k), \\[30pt]
    0, & i\notin \mathcal S(k),
    \end{cases}
\end{equation}
with $\hat{k}_a$ being the wavenumber associated with a node/bin, and the index $m$ running over the nodes within the set $\mathcal{S}(k)$. The latter set comprises the four closest nodes, $k_a$, to the input value $k$. For a given $k$, we define the set $\mathcal{S}(k)$. If node $i$ is not within the set then we return $\psi_i = 0$. If it is within the set, then the polynomial depends on the value of $i$. The term $\psi_i(k)$ is simply one way of smoothly interpolating between values across $N$ nodes, and it is defined by the constraint $\psi_i(\hat{k}_i) = 1, \psi_i(\hat{k}_{m \neq i}) = 0$, i.e. the interpolator must recover the input value at its own bin centers and not alter the evaluated value at other bin centers.

In practice, we use a different kernel for the external, $k_3$ leg. This kernel is denoted by $\kappa_{ijl}$ and is distinguished by its dependence on all three indices. That is, the specific basis function we use depends on the chosen bins/nodes of $\kOne$ and $\kTwo$. The kernel can then be written as,
\begin{equation}\label{eqn:kernel:cubic_cellbased}
    K_{12}(\kOne,\kTwo,\kThree) \approx \sum_{i j l} \psi_i(\kOne) \psi_j(\kTwo) \kappa_{ijl}(\kThree),
\end{equation}
where, as we will show below, we have absorbed the coefficient $\alpha_{ijl}$ into the definition of $\kappa_{ijl}$. Note that this does not break the separability ansatz given $\kOne$ and $\kTwo$ only determine the choice of basis function for $\kThree$. The term we evaluate is still $\kappa_{ijl} (\kThree)$, which only has $\kThree$ as its functional argument and is not dependent on $\kOne, \kTwo$ as is required for the ansatz to hold. The functional form of $\kappa_{ijl} (\kThree)$ is a simple cubic polynomial,
\begin{equation}
    \kappa_{ijl}(k)
    =
    \Theta_l(k)
    \left[
    \alpha_{ijl}^{(0)}
    + \alpha_{ijl}^{(1)} u_l(k)
    + \alpha_{ijl}^{(2)} u_l(k)^2
    + \alpha_{ijl}^{(3)} u_l(k)^3
    \right] .
\end{equation}
where the local coordinates are defined as,
\begin{equation}
    u_l(k)
    =
    \frac{k - \hat{k}_l}
         {\tilde{k}_{l, +} - \tilde{k}_{l, -}} ,
\end{equation}
with $\tilde{k}_{l,+}$ and $\tilde{k}_{l,-}$ denoting the right and left edges of a given bin, and $\hat{k}_l$ denoting the center of the same bin. The coefficients, $\alpha_{ijl}^{(a)}$, are determined by fitting the input kernel function at the bin center and the edges. Fitting to the edges ensures continuity in the polynomials across different cells. For similar reasons, we also evaluate the derivative of the kernel with respect to $\kThree$ (using finite differencing\footnote{While finite differences can suffer from numerical noise, we find our results are stable to varying the choice of width in the numerical derivatives. Thus, numerical noise is inconsequential for the setup in this work.}) and enforce that it is also matched by the polynomial. By definition, $\alpha_{ijl}^{(0)}$ is the value of the kernel evaluated at the bin center.

The final kernel decomposition is given as,
\begin{equation}\label{eqn:kernel:final}
    K_{12}(\kOne,\kTwo,\kThree) = \sum_{i j l} \psi_i(\kOne) \psi_j(\kTwo) \Theta_l(\kThree)
    \left[
    \alpha_{ijl}^{(0)}
    + \alpha_{ijl}^{(1)} u_l(\kThree)
    + \alpha_{ijl}^{(2)} u_l(\kThree)^2
    + \alpha_{ijl}^{(3)} u_l(\kThree)^3
    \right].
\end{equation}
All results in our work utilize this choice of kernel decomposition. The FFT-based integration in Equation \eqref{eqn:sim:ICs:S12} can be rewritten as,
\begin{align}\label{eqn:sim:ICs:Final}
    \phi_{\rm NG}(\vec{k}) = &\,\, \phi_{\rm G}(\vec{k}) +  \fNL \sum_{i j l}
    \bigg[\Theta_l(\kThree)\left(
    \alpha_{ijl}^{(0)}
    + \alpha_{ijl}^{(1)} u_l(\kThree)
    + \alpha_{ijl}^{(2)} u_l(\kThree)^2
    + \alpha_{ijl}^{(3)} u_l(\kThree)^3
    \right) \times  \nonumber\\
    & \hspace{90pt}\FFT\Big\{\iFFT \Big(\psi_i(\kOne) \phi_G(k_1)\Big) \times \iFFT \Big(\psi_j(\kTwo)\phi_G(k_2)\Big)\!\Big\}\!\bigg],
\end{align}
This is a fully separable approach but does not assume any global basis functions. The only functions we have utilized are different interpolating kernels, whose purpose is to effectively smooth over the binned estimates.

There is, however, one more decision to be made: the construction of a kernel, $K_{12}$, for a given model. This is particularly relevant because there are multiple, equivalent kernels that can be constructed from a given bispectrum model; the availability of choice arises from the permutation symmetries in the input wavenumbers. However, while the different choices of kernels all produce the same bispectrum, they can have different infrared (IR) corrections to the primordial power spectrum. In many cases the corrections can be divergent, causing it to be larger than the original ``tree-level'' primordial power spectrum. \citet{Scoccimarro2012PNGs} discuss this in detail and introduce strategies for constructing linear combinations of the different, equivalent kernels such that the IR divergences are canceled. In \citetalias{paper1} and \citetalias{paper2}, we extended this approach for generic basis functions, and applied it to over a hundred different input bispectra. In all cases, the IR divergence was handled by carefully choosing the functional form used in approximating the kernel.

In the case of a binning approach like the one used in this work, there is no choice of functional form,\footnote{If we enforced some functional form, the binned approach would simply reduce back to a basis decomposition approach, such as that from \citetalias{paper1}.} so we must find a generic method for reducing IR divergences. For this, we rely on the methodology of \citet{Wagner:2011:PNG}, who use the reduced bispectrum as their kernel,
\begin{equation}\label{eqn:k12_reduced}
    K_{12}(\kOne, \kTwo, \kThree) = \frac{\Bppp(\kOne, \kTwo, \kThree)}{
    P_{\phi \phi}(\kOne) P_{\phi \phi}(\kTwo) + P_{\phi \phi}(\kTwo) P_{\phi \phi}(\kThree) + P_{\phi \phi}(\kOne) P_{\phi \phi}(\kThree)}
\end{equation}
where $P_{\phi \phi}$ and $\Bppp$ are the power spectrum and bispectrum of the inflaton potential field. Under the choice of kernel above, the behavior in the IR limit ($k\rightarrow0$) is well controlled as has been explicitly demonstrated by \citet{Fondi:2025:Gengars}. The choice in Equation \eqref{eqn:k12_reduced} is generally challenging to use as even in the case of a separable bispectrum, the method constructs a non-separable kernel. \citet{Fondi:2025:Gengars} handled this by using a Schwinger-approximation, at the cost of increasing computational runtime by a factor of $150$. In our case, we are already approximating non-separable kernels/bispectra and thus do not incur extra difficulties if we construct the kernel in this manner. Thus, Equation \eqref{eqn:k12_reduced} is the approach we use in this work. We have checked that corrections to the power spectrum have an amplitude $d\ln P / d\fNL \lesssim 10^{-5}$, which is completely negligible for our analysis.

\begin{figure}
    \centering
    \includegraphics[width=\columnwidth]{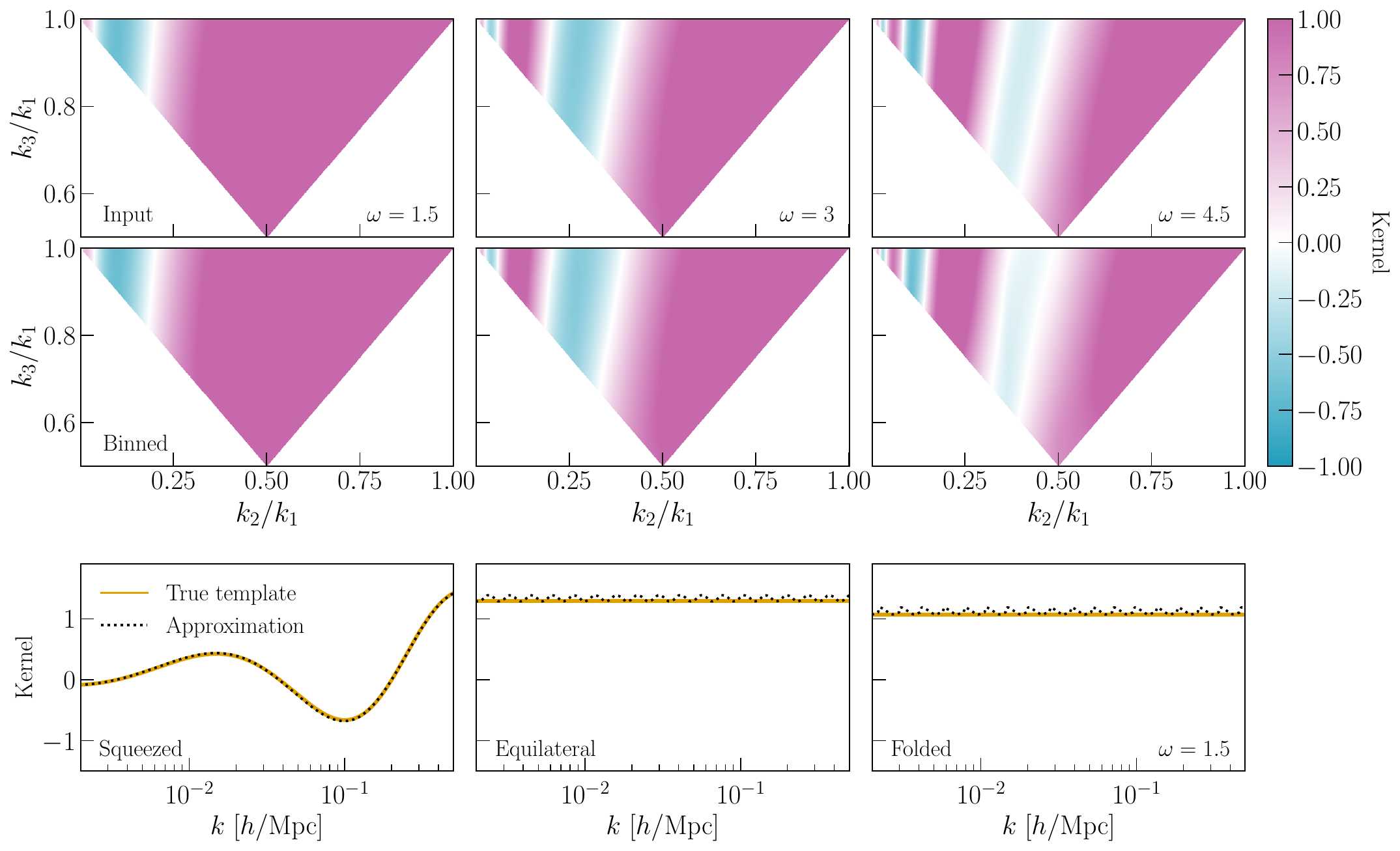}
    \caption{\textit{Upper:} The kernel function for the QSFosc model of Equation \eqref{eqn:model:QSFosc}, plotted in different kinematic limits for three different frequencies (columns). The left, bottom, and right corners of the triangle represent the squeezed, folded, and equilateral limits, respectively. The color bar shows the value of the kernel. We see oscillations towards the squeezed limit that do not extend towards the folded and equilateral limits given the truncation used in defining the model. The binned approximation (bottom) reproduces the features of the input kernel (top). \textit{Bottom:} The three limits of the kernel function for the $\omega = 1.5$ model. Our approximation can precisely match the non-monotonic behavior in the squeezed limit while also representing the other two limits accurately. The residual oscillations in the other limits arise due to cubic interpolation over the coarsely binned function; their amplitudes are at the few-percent level. For brevity, we show this only for the $\omega = 1.5$ model that is the main focus of our work.}
    \label{fig:ICmethod}
\end{figure}

Figure \ref{fig:ICmethod} shows an example of the collider models that we describe in Section \ref{sec:sims:cosmocollider} below, alongside their approximated versions. We use 20 bins in each axis, spaced logarithmically in $k$. We can accurately reproduce the features of the bispectrum across the whole function space. The bottom panels show the kernel in specific limits: (i) squeezed limit, $\kThree \ll \kOne \approx \kTwo$, (ii) equilateral limit, $\kOne \approx \kTwo \approx \kThree$, and (iii) folded limit, $\kTwo + \kThree \approx \kOne$. We can precisely reproduce the behavior of the kernel in these specific limits. There are residual oscillations in the equilateral and folded limits, arising from the cubic interpolation kernel we use. However, we have compared halo bias predictions from the true bispectrum to those from the approximated one and find they agree to better than 1\%. So these few percent-level residuals are negligible for the purposes of our work here.

\subsection{Cosmological Collider models}\label{sec:sims:cosmocollider}

There are a wide variety of cosmological collider signatures, each arising from specific choices for the additional field(s) present during inflation \cite[\eg][]{Nima:2015:Colliders, Dizgah2020CosmoCollider, Sohn:2024:Colliders}. In this work, we focus on the generic predictions from these models; namely that in the squeezed limit, the primordial bispectra scale as $B(k,k, q) \propto (k/q)^\Delta \cos(\log(k/q))$. We start from the template for a quasi-single field model, which captures the impact of massive (but still relatively light) particles with $m/H_\phi \leq 3/2$. Here, $m$ is the mass of the additional field and $H_\phi$ is the Hubble rate during inflation. The resulting collider model has no oscillations and is generated by a cubic interaction, $\sigma^3$, in the heavy sector together with linear mixing, $\dot\phi\sigma$. We use the standard template defined by~\citep{Chen:2010:QSF},
\begin{equation}\label{eqn:model:QSF}
    S^{\rm QSF} = 3\sqrt{3k_t} \,\frac{Y_\nu(8k_t)}{Y_{\nu}({8}/{27})}\,,   
\end{equation}

\noindent where $Y_\nu$ is the Bessel function of the second kind, $k_t = \frac{\kOne\kTwo\kThree}{(\kOne + \kTwo + \kThree)^3}$, $\nu = \sqrt{9/4 - (m/H_\phi)^2}$ and $S(\kOne, \kTwo, \kThree) = (\kOne\kTwo\kThree)^2 B(\kOne, \kTwo, \kThree)$ is a shape function. In this work, we consider $\nu = 1$ which results in a squeezed-limit scaling $(k/q)^{0.5}$ for the template, and $k^{-1.5}$ for the halo bias. Next, we apply the oscillations in only the squeezed limit, as
\begin{equation}\label{eqn:model:QSFosc}
    S_{\rm QSFosc} = S_{\rm QSF} \times \bigg[1 + 4 \cos\bigg(\omega \ln \frac{\kOne}{\kThree} + \phi\bigg)\Theta(\kThree/\kOne < 0.15)\bigg] + {\rm perms.}
\end{equation}
where the factor of 4 simply boosts the oscillation signal. We will consider models with $\omega = \{1.5, 3, 4.5\}$, but focus primarily on $\omega = 1.5$ for brevity. The Heaviside function $\Theta(\kThree/\kOne < 0.15)$ in Equation \eqref{eqn:model:QSFosc} ensures the oscillations are only present in the squeezed limit and completely removed from the folded and equilateral limits. The cosmological collider models predict oscillations predominantly in the former limit. Thus, extending the same oscillatory structure to the latter two limits would be unphysical and testing model phenomenologies outside our interest. Thus, we follow prior choices, such as in \citet{Dizgah2020CosmoCollider}, and apply step functions to remove the oscillations outside the squeezed limit. This is also one of the key choices that distinguishes our work from \citet{Goldstein:2025:CosmoColl}, who also performed a simulation study of the collider signal but limited their initial conditions construction and subsequent analysis to the squeezed-limit of the bispectrum.\footnote{This was because they could construct a reliable, separable kernel for the squeezed-limit behavior alone, and therefore focused their analysis of the simulations on just that limit.}

While the $S_{\rm QSFosc}$ model is not tied to an inflationary Lagrangian in the early Universe, we prefer it over more fundamental models, as it is easier to identify the key parameters impacting the phenomenological features (such as the frequency and phase of the oscillations). This also allows us to avoid the oscillations being exponentially suppressed at higher frequencies, which is a generic feature of such models \citep{Nima:2015:Colliders} and would make it challenging to precisely measure the signals considered in this work. However, \citet{Oliver:2026:ColliderLimits} note that the suppression can be alleviated by considering stronger mixing between the inflaton and the additional field. To summarize, we stress that our results generalize to any collider model as we include the physically motivated, oscillatory signal in the squeezed limit, but the choice of $S_{\rm QSFosc}$ enables us to make precision measurements using a smaller set of simulations than would otherwise be necessary if the signal was exponentially suppressed.

\subsection{Simulations and halo bias measurements}\label{sec:sims:simconfig}

Using the methodology of Section \ref{sec:sims:ICs}, we generate initial conditions for the model described in Section \ref{sec:sims:cosmocollider} and produce a set of N-body simulations. The latter is done using the \textsc{Pkdgrav3} N-body code \citep{Potter2017Pkdgrav3}, which was also used (and tested) in our previous explorations of PNGs \citep[][\citetalias{paper1}, \citetalias{paper2}]{Anbajagane2023Inflation}. We produce 10 simulations (per model) to reduce cosmic variance in our halo bias measurements. Given the signatures of interest are on larger scales, we use a box size of $L = 3 \gpc/h$, which has a fundamental mode $k_F = 2\pi/L = 2 \times 10^{-3} h/\mpc$. All simulations are run with $1024^3$ dark matter particles, and have a particle mass of $2 \times 10^{12} \msun/h$. Therefore, halos of $M = 10^{14} \msun/h$ are resolved with 50 particles. For the purposes of this work, where we are interested in the large-scale clustering statistics of collapsed structures, low resolution impacts our measurements via the selection function of the simulated halo sample. 

At high resolution, all halos of a given mass are identified by the halo-finding procedure; the completeness of halo finding is essentially $f = 1$. At the low-resolution limit, the halo finder can preferentially find more/less concentrated halos depending on the finder's configuration \citep[\eg][]{Woodring:2011:Concentration, More:2011:SO}. As a result, there is a concentration-based selection in the sample of the most poorly resolved halos. This can be trivially accounted for in our modeling approach, as we discuss further in Section \ref{sec:results}, because these low-resolution objects are still collapsed structures that are biased tracers of the density field. Given our modeling can handle their properties, we still extend our analysis down to halos of $M \approx 10^{14} \msun/h$. No modeling subtleties are observed in our analysis of halos with $M \geq 2 \times 10^{14} \msun/h$.

Our simulations output ten snapshots spanning $0 < z < 10$,\footnote{Note that this has no bearing on the time-stepping criterion used in the differential equation solvers. The snapshot count only sets how many data products are output/stored.} and we run the \textsc{Rockstar} halo finder \citep{Behroozi2013Rockstar} on each of the ten snapshots. Throughout this work, we will use the definitions $\Mtwohc$ and $\Rtwohc$ for the halo mass and radius. The latter is the halo-centric distance within which the average density is 200 times the critical density at that epoch, while the former is the mass contained within that radius. \textsc{Rockstar} also measures the maximum circular velocity, $\Vmax$, of the mass distribution within the halo. This is a proxy for the halo concentration and is generally more reliable \citep[\eg][]{Prada:2012:Concentration} as it does not require measuring and fitting a finely binned density profile within the halo. 

We will use the normalized quantity, $\tVmax = \Vmax / \Vtwohc$ with $\Vtwohc = \sqrt{G \Mtwohc / \Rtwohc}$ being the characteristic velocity scale of the halo. We have confirmed that the mean $\tVmax-\Mtwohc$ relation in our simulations matches those found in the literature \citep[\eg][]{Prada:2012:Concentration, Mansfield:2021:Biased}, and the log scatter of $0.1 {\rm\,\, dex}$ is also consistent with the literature. Though our simulations do not have the resolution to do detailed studies of halo internal structure, the $\Vmax$ estimates are stable enough to perform the coarse two-quantile binning in Section \ref{sec:results:assembly}.

Using these simulations, we measure the halo bias,
\begin{equation}
    b(k) = P_{\rm hm}(k)/P_{\rm mm}(k)
\end{equation}
as the ratio of the halo-matter cross spectrum to the matter auto spectrum. This ratio is cosmic-variance (CV) suppressed, which is vital as our measurements are CV-dominated at the larger scales we measure the bias on. We measure both power spectra on grids of $N_{\rm grid} = 512$. The halo field is obtained by simply counting the number of halos (after a given selection is applied) that fall into a cell in the grid. The overdensity field is obtained by similarly counting dark matter particles on the grid and performing the transform, $\delta = N / \langle N \rangle - 1$, where $N$ is the particle-count field.

\subsection{Theoretical predictions for halo bias}\label{sec:sims:theory}

The introduction of the primordial bispectrum, $\Bppp$, alters the formation of collapsed structure in a spatially dependent way. We want to compute this scale-dependent correction, $\Delta b (k)$. This can be done as in \citep{Desjacques:2018:Bias},
\begin{equation}\label{eqn:theory:delta_b}
    \Delta b(k, M, z)
    =
    2f_{\rm NL}
    \frac{\mathcal{F}_R(k, M, z)}{\mathcal{M}(k,z)}
    \left[
        b_\phi(M, z)\, 
        +
        2b_\sigma\, \frac{\partial \ln \mathcal{F}_R(k, M, z)}{\partial \ln \sigma_R^2}
    \right],
\end{equation}
where the second term in the brackets accounts for the impact of PNGs on the Jacobian, i.e. the remapping of a mass interval $d\ln M$ to a peak-height interval, $d\ln \nu$.\footnote{The peak height is defined as $\delta_c / \sigma_R$ with $\delta_c = 1.686$ and $\sigma_R$ given by Equation \eqref{eqn:sigmaR}.} The coefficient $b_\phi(M, z)$ is the well-known response of halo number density to modulations in the large-scale potential modes and approximately behaves as $b_\phi = \delta_c (b_1 - 1)$ for a halo mass-selected sample \citep{Dalal2008ScaleDependentBias}. Here, $b_1$ is the scale-independent linear bias term. The coefficient $b_\sigma$ represents the response of the abundance to a local distortion in the $dM-d\nu$ remapping, and takes a value $b_\sigma = 1$ for a purely mass-selected sample. Both coefficients deviate from their default values when halo samples are constructed from more complicated selections.\footnote{Performing selections on secondary halo properties, such as halo concentration, breaks the universality assumptions of the halo mass function. As a result, the coefficients above take corrections that depend on the exact selection.} 

The function $\mathcal{F}_R$ is obtained by integrating over the primordial bispectrum,
\begin{equation}\label{eqn:Fr}
    \mathcal{F}_R(k, M, z)
    =
    \frac{1}{4 \sigma_R^2 P_\Phi(k)}
    \iint \frac{q^2 dq\, d\mu}{(2\pi)^2}\,
    \mathcal{M}_R(q,z)\,
    \mathcal{M}_R(p,z)\,
    \Bppp(k,q,p),
\end{equation}
with the integral evaluated over $q \in [10^{-4}, 20]$ and $\mu \in [-1, 1]$. The internal momentum, $p$, is defined as $p \equiv \sqrt{k^2 + q^2 - 2 k q \mu}$, and the transfer kernel is defined as
\begin{align}
    \mathcal{M}_R(k,z) & \equiv \mathcal{M}(k,z) W_R(k),\nonumber\\[10pt]
    W_R(k) & \equiv \frac{3 j_1(kR)}{kR},
\end{align}
This kernel includes both the standard transfer function $\mathcal{M}(k,z)$ --- which converts between the primordial potential and density modes as $\delta_m(\mathbf{k},z)  = \mathcal{M}(k,z)\Phi(\mathbf{k})$; we obtain this from \textsc{CCL} \citep{Chisari2019CCL} --- and the smoothing term corresponding to a halo of Lagrangian size, $R$. The latter radius is defined as,
\begin{equation}
    R(M)
    =
    \left(
        \frac{3M}{4\pi \bar{\rho}_{m,0}}
    \right)^{1/3},
\end{equation}
where $\bar{\rho}_{m, 0}$ is the comoving mean matter density at the present epoch. The variance term in Equation \eqref{eqn:Fr} is given as
\begin{equation}\label{eqn:sigmaR}
    \sigma_R^2
    =
    \int \frac{q^2 dq}{2\pi^2}\,
    \mathcal{M}_R^2(q,z)\,
    P_\Phi(q).
\end{equation}

We can inspect Equation \eqref{eqn:theory:delta_b} to better understand its phenomenology. The first term in the brackets simply rescales the amplitude by some amount. The second term has an additional, more non-trivial mass dependence. We can rewrite the derivative as, 
\begin{equation}
    \frac{\partial \ln \mathcal{F}_R(k, M, z)}{\partial \ln \sigma_R^2} = \frac{\partial \ln \mathcal{F}_R(k, M, z)}{\partial M}\frac{\partial M}{\partial \ln \sigma_R^2}.
\end{equation}
Therefore, the second term contains the derivative of the kernel with mass. For bispectra that contain oscillations via a cosine term, the derivative will include a sine term. Thus, Equation \eqref{eqn:theory:delta_b} contains a linear combination of cosine and sine terms, which together generate a phase offset in the oscillations. As we will discuss in Section \ref{sec:results}, these offsets are indeed measured and are well fit by the model described above.

For the model originally studied in \citet{Dalal2008ScaleDependentBias} --- henceforth referred to as Local-type PNG --- we see $\mathcal{F}_R = 1$, as the integral in Equation \eqref{eqn:Fr} evaluates to just $\sigma_R^2$ and cancels with the prefactor. Consequently, most literature on the Local-type PNG model omits this term and only include the transfer function $\mathcal{M}$. The fact that $\mathcal{F}_R = 1$ also means the derivative $\partial \mathcal{F}_R / \partial X = 0$ (for any variable $X$) and thus the second term in Equation \eqref{eqn:theory:delta_b} is omitted in discussions of this model. Both terms are necessary for most other models, including the ones used in this work.

\section{Results}\label{sec:results}

We now present the measurements from our simulations for various collider signatures and halo sample definitions. We then discuss the theoretical model's flexibility in fitting these measurements.

The fiducial simulations we use follow Equation \eqref{eqn:model:QSFosc} with $\omega = 1.5$ and $\phi = 0.5$. As we will show later, the former choice is optimal for increasing the amplitude of the measurable feature as higher frequencies suppress the amplitude of the signal in the halo bias. We use $\phi = 0.5$ as it shifts the peak of the oscillation into a $k$-mode range that is sufficiently linear while also being reliably measured in our simulations. We discuss higher-frequency variants with $\omega \in \{3, 4.5\}$ to showcase the suppression of the oscillations' amplitudes. In all cases, we will present the CV-suppressed ratio, $b^+/b_1$, where $b^+$ is the bias from simulations with $\fNL = 500$ and $b_1$ that from the Gaussian run with $\fNL = 0$. We will show the mean ratio with error bars obtained from bootstrap resampling over the 10 simulation realizations. All results are presented at $z = 0$. At higher redshifts, we see similar phenomenology as what we discuss below so we focus our discussions on just $z = 0$ for brevity.

\subsection{Oscillating halo bias and mass-dependent phases}\label{sec:results:bias}

\begin{figure}
    \centering
    \includegraphics[width=\columnwidth]{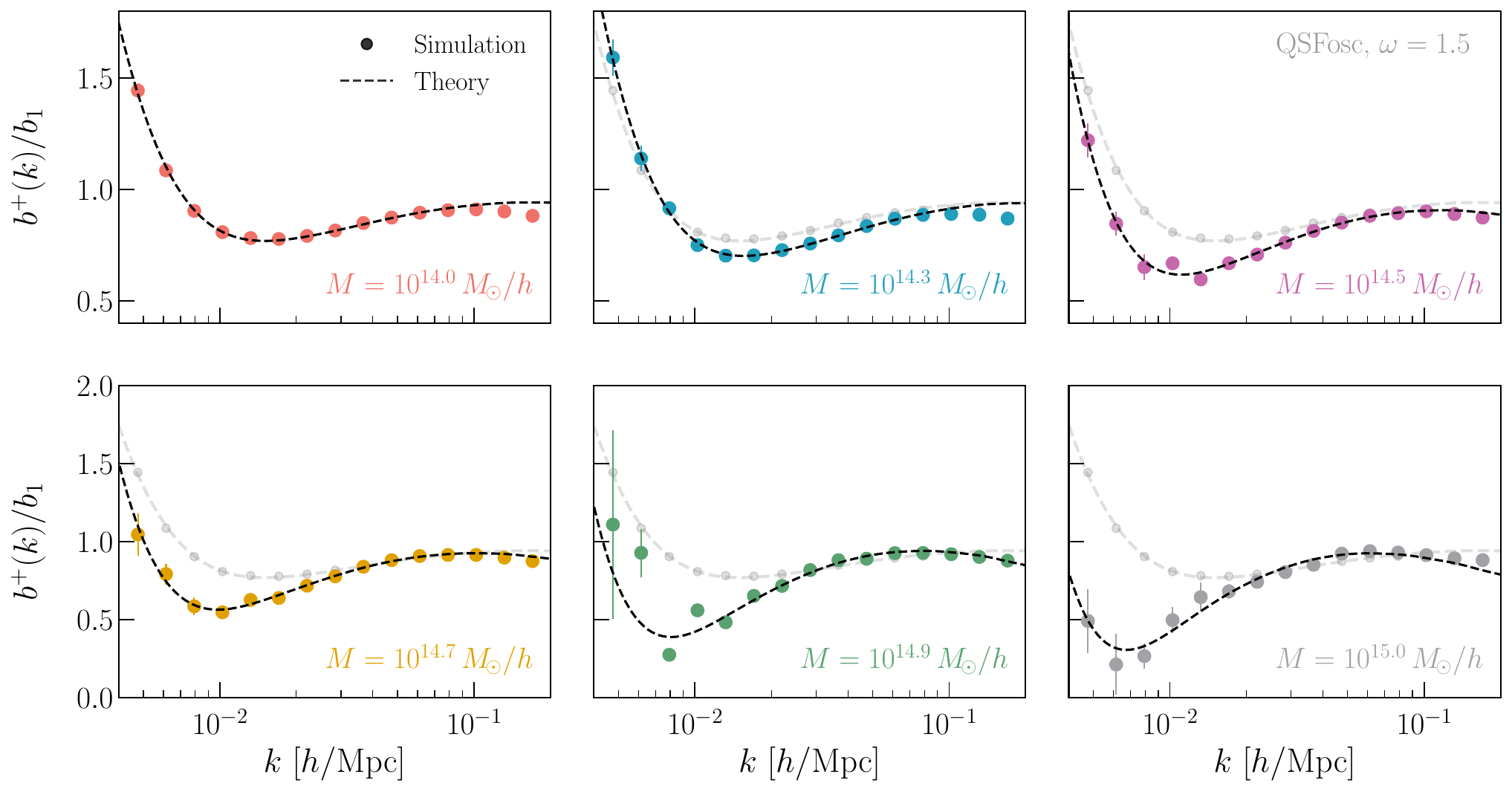}
    \caption{The halo bias measurements from simulations (points) alongside theory predictions (dashed lines) for different halo mass bins of width $\Delta M = 10^{14} \msun/h$. We replicate the lowest-mass bin's results as small gray points and lines in the rest of the panels, to visually highlight the evolution of the PNG signature with halo mass. There is a clear presence of an oscillation originating from the primordial bispectrum, and the feature's amplitude and location are strongly mass-dependent.}
    \label{fig:OscBias}
\end{figure}

Figure \ref{fig:OscBias} shows the key result of this paper --- a scale-dependent, oscillating halo bias in simulations of cosmological collider models. The bias ratio displays a clear non-monotonic behavior as a function of scale, $k$. The ratio is less than 1 on intermediate scales and increases towards larger scales.\footnote{Note that the oscillations will continue towards even larger scales if the halo-bias measurements could be extended further. On such scales, the asymptotic scaling of $k^{-1.5}$ dominates the behavior, but oscillations will still modulate this scaling.} Each panel in Figure \ref{fig:OscBias} replicates the measurement and theory prediction from the lowest-mass bin as small gray points and a thin gray line, respectively, for visual reference. We can see a clear difference between the bias exhibited by the different mass selections. The characteristic scale, $k_c$, where the oscillation amplitude is largest, is mass-dependent. Larger halos clearly exhibit smaller $k_c$ values (larger scales). This is due to the mass-based kernel in Equation \eqref{eqn:Fr}. Changing the halo mass by a factor of ten results in the halo radius $R \propto M^{1/3}$ changing by a factor of $2.15$. Unsurprisingly, we find the scale $k_c$ is shifted by a similar factor, from $k_c = 0.015 \rightarrow 0.0068 \,\,h/\mpc$, across the panels.\footnote{We compute $k_c$ by fitting a cubic spline and determining the location of the first trough.} The amplitude is also mass-dependent, and increases as we focus on higher mass objects. This may seem obvious given the PNG feature scales with $b_\phi \sim b_1 - 1$ and the ratio of $b_\phi / b_1$ increases a little with mass. However, it is also dependent on the mass-dependent kernel, $W_R$, used in Equation \eqref{eqn:Fr}. Just rescaling the lowest-mass template by a larger $b_\phi$ does not reproduce the amplitude of the highest-mass template.

The dashed lines in Figure \ref{fig:OscBias} demonstrate that the theory model from Equation \eqref{eqn:theory:delta_b} provides a precise fit to the simulation measurements. We find the best-fit values of $b_\phi$ and $b_\sigma$ match their corresponding universality predictions (Appendix \ref{appx:bias}). There is a slight deviation for the lowest-mass bin, where halos are poorly resolved given our particle mass and are therefore not a purely mass-selected sample but rather a sample with some concentration-based selection. All other bins do not exhibit such behavior and their coefficients are consistent with universality.

\subsection{Assembly bias}\label{sec:results:assembly}

In Section \ref{sec:results:bias} above, we noted how the values of $b_\phi$ and $b_\sigma$ match their universality-based predictions; i.e., $b_\phi = \delta_c (b_1 - 1)$ and $b_\sigma = 1$. For this ansatz to hold, the sample being studied must be selected only by halo mass. Any additional selection, which cannot be represented purely as a selection in mass, will break this ansatz and result in the parameters taking different (a priori unknown) values. We now explicitly test this by splitting our halo sample based on $\tVmax$, which is a proxy for the halo concentration. We compute the median value of $\tVmax$ in each halo bin and split the sample based on this value. Because our halos have between 50 and 500 particles, the measurements of $\tVmax$ will be impacted by limited resolution. However, since we only separate halos into two quantiles, the requirements on precision are significantly loosened; any numerical scatter smaller than the bin width will have little impact on the final results, and our bin widths are large given we only use two quantiles. We have also checked that the $\tVmax - \Mtwohc$ relation qualitatively matches results from the literature \citep[\eg][]{Prada:2012:Concentration, Mansfield:2021:Biased}.

\begin{figure}
    \centering
    \includegraphics[width=\columnwidth]{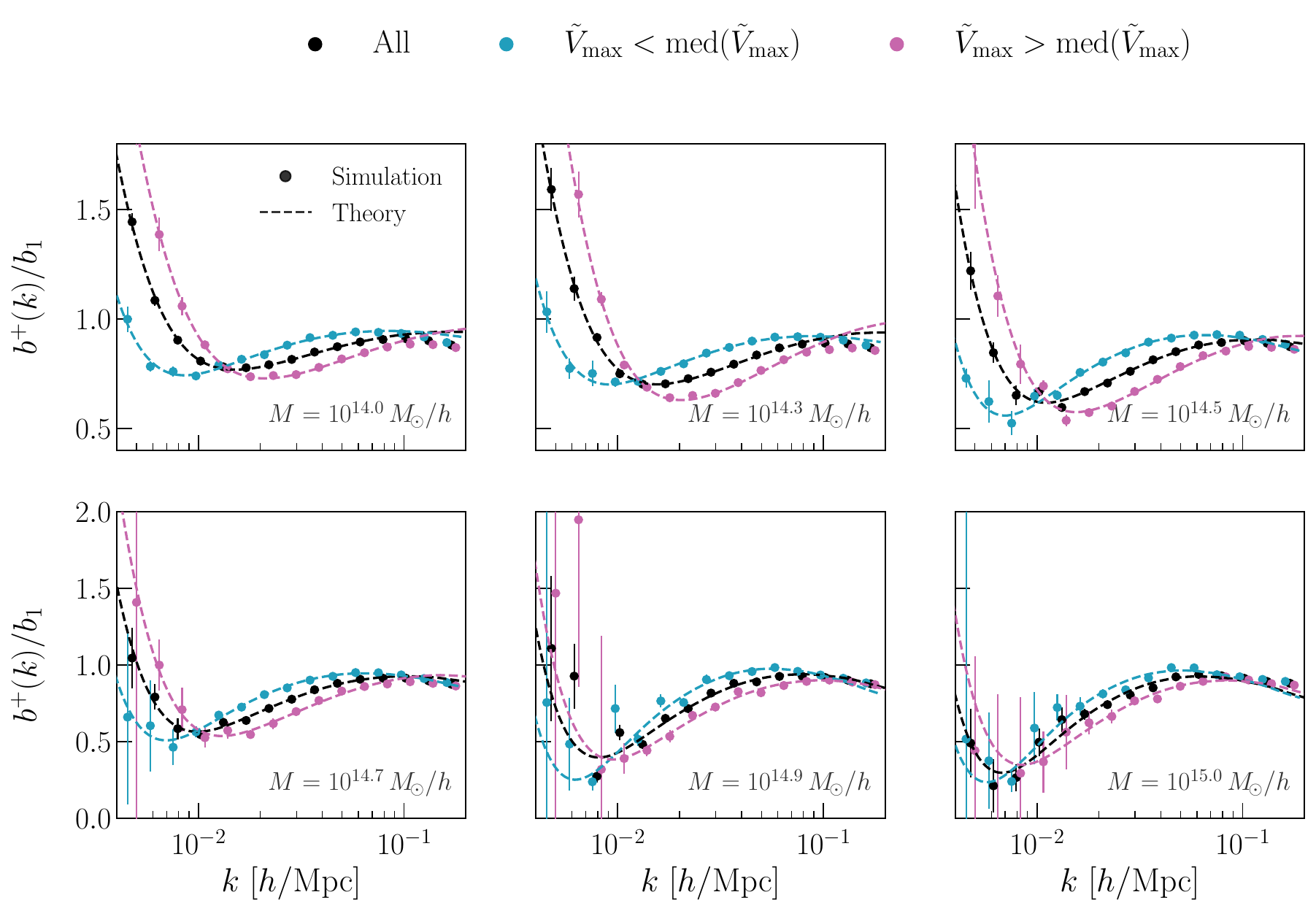}
    \caption{The effect of assembly bias on the measurements of scale-dependent halo bias. We see a clear shift in the phase of the oscillating term that depends on $\tVmax$. More (less) concentrated halos exhibit features on  smaller (larger) scales than the fiducial sample. Our two-parameter model in Equation \eqref{eqn:theory:delta_b} provides an adequate fit to the simulation measurements, though the $b_\phi$ and $b_\sigma$ values correlate strongly with $\tVmax$ selection (see Figure \ref{fig:bphi_bsigma}).}
    \label{fig:OscBias_Vmax}
\end{figure}

Figure \ref{fig:OscBias_Vmax} shows measurements of the scale-dependent bias from $\tVmax$-binned halo samples. In all mass bins, there are clear offsets in the phase of the halo bias oscillations across the different $\tVmax$ selections. For a sample with the same mean halo mass, the location of the oscillations, $k_c$, can shift by as much as a factor of two due to assembly bias. Notably, this phase shift is similar in size to the mass-dependent shift we found across an order of magnitude in halo mass, discussed above. Specifically, we find the wavenumber $k_c$ to shift by a factor of 2 at $10^{14} \msun/h$ masses, and a factor of 1.2 at $10^{15} \msun/h$ masses. The relative amplitude of these shifts follows closely from the scatter in $\tVmax$, which we find is 1.6 times larger at $10^{14} \msun/h$ relative to that at $10^{15} \msun/h$.\footnote{While the scatter in the halo concentration --- which is the target quantity that $\tVmax$ serves as a proxy for --- has a nearly constant scatter down to $M \approx 10^{11} \msun/h$ \citep[\eg][]{Anbajagane2022Baryons}, the derivative between concentration and $\tVmax$ is larger for halos of lower masses \citep{Prada:2012:Concentration}, as such halos have higher concentrations on average. So, a nearly constant scatter in concentration is converted to a mass-dependent scatter in $\tVmax$.} We determined these values using the same cubic-spline procedure noted in Section \ref{sec:results:bias}. Our results from the higher frequency models, presented in Section \ref{sec:results:higher_freq} below, show the size of this $\tVmax$-based shift decreases with increasing $\omega$.

The above results confirm that for an observed sample of biased tracers, the selection on secondary halo properties can be just as important as the mass-based selection. Figure \ref{fig:OscBias_Vmax} also demonstrates that the model in Equation \eqref{eqn:theory:delta_b} captures the behavior for different halo selections. That is, the freedom provided by the parameters $b_\phi$ and $b_\sigma$ is adequate for fitting the simulated measurements. The best-fit values of these parameters are detailed further in Appendix \ref{appx:bias}.

\subsection{Oscillation amplitudes in higher-frequency models}\label{sec:results:higher_freq}

\begin{figure}
    \centering
    \includegraphics[width=\columnwidth]{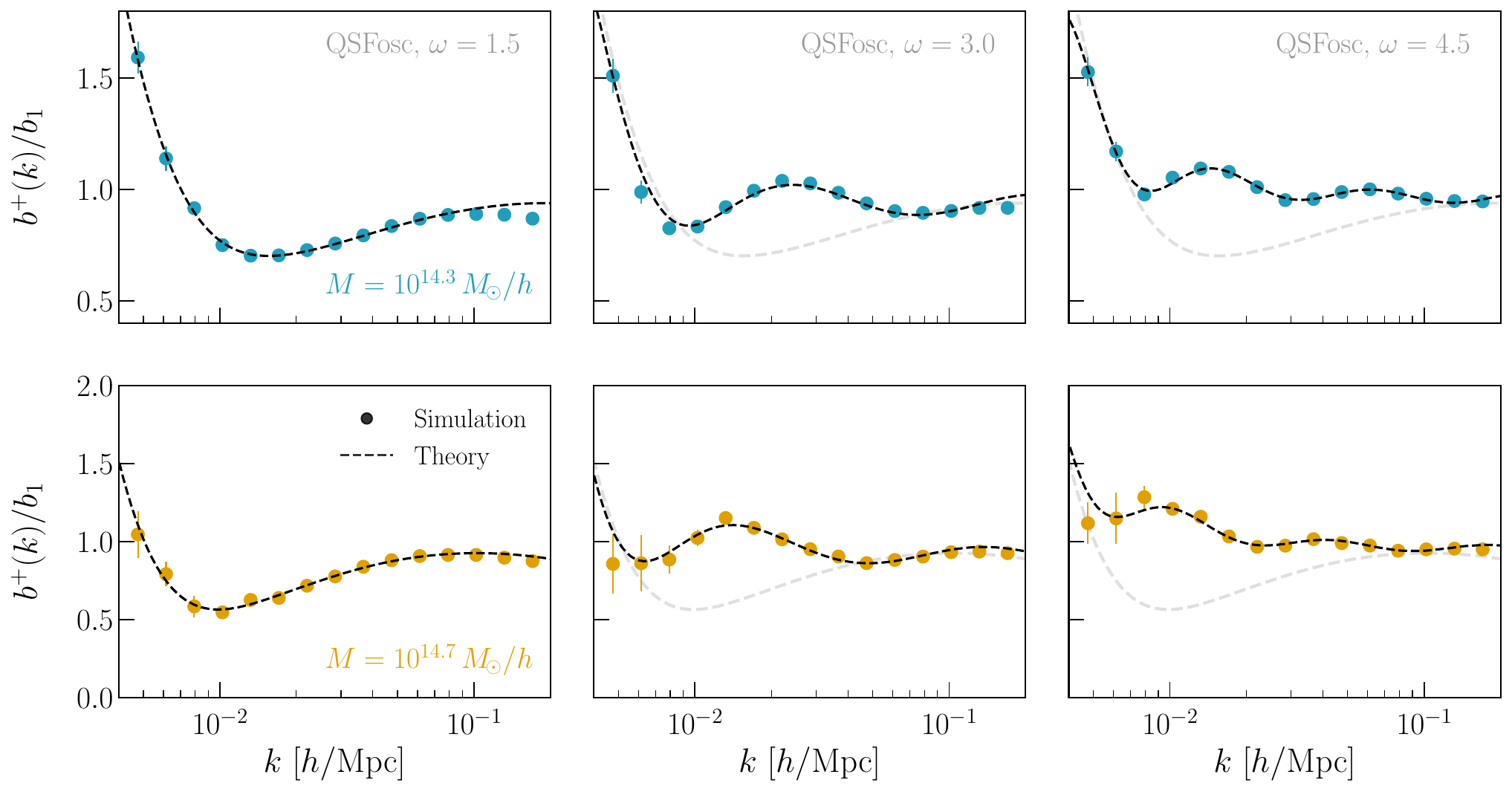}
    \caption{Similar to Figure \ref{fig:OscBias}, but now for models at three different frequencies and for just two of the six mass bins. We can accurately measure and predict the oscillations for higher frequencies, $\omega$. The amplitude of the oscillations decreases with increasing frequency, as the oscillations can more easily average down within the window function of a halo (see Figure \ref{fig:OscBias_freq_scan}). The theory prediction for $\omega = 1.5$  (gray line) is shown in the other two columns as a reference.}
    \label{fig:OscBias_combined}
\end{figure}

\begin{figure}
    \centering
    \includegraphics[width=\columnwidth]{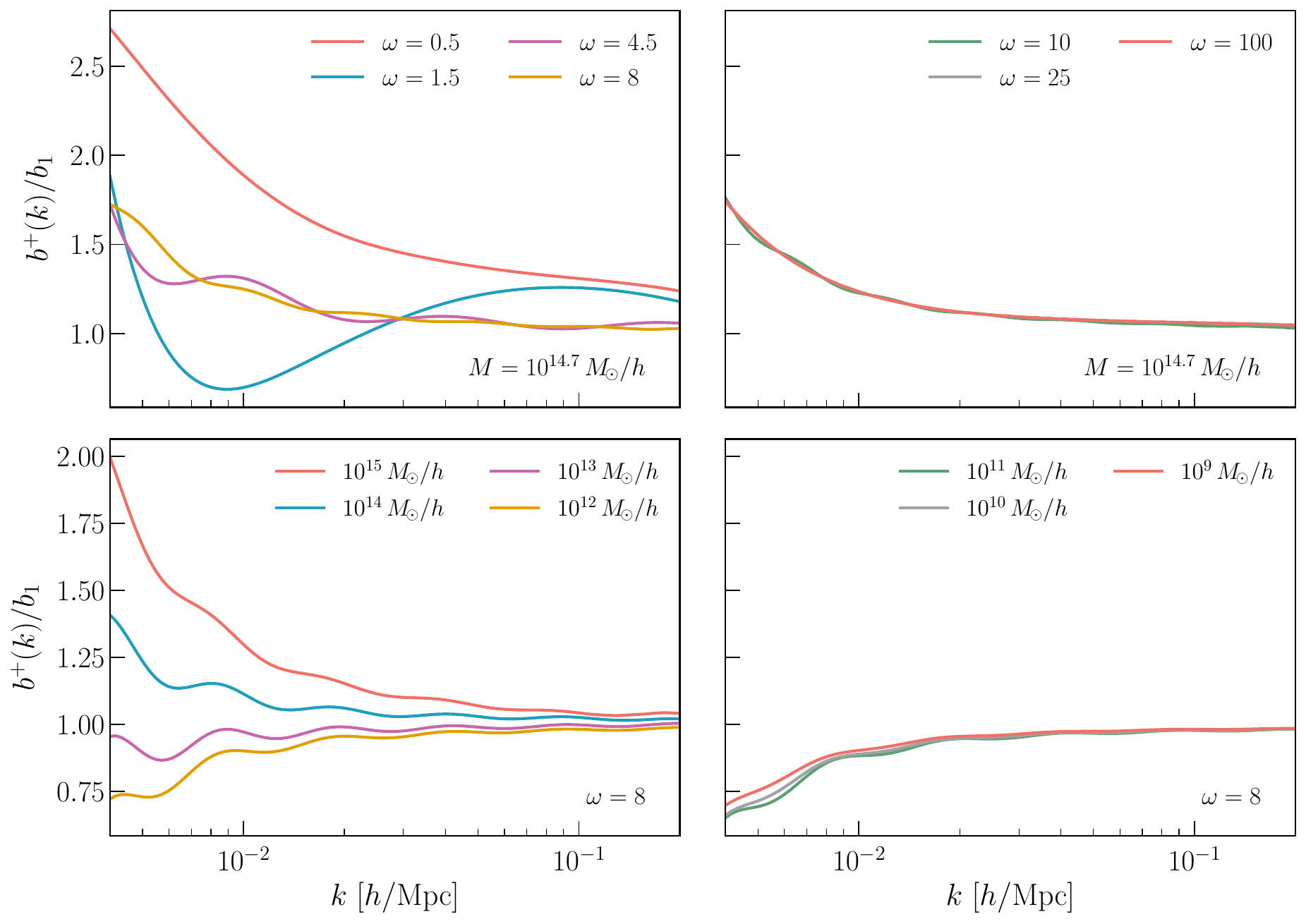}
    \caption{\textit{Top:} the theory predictions for the halo bias at $M = 10^{14.7} \msun/h$ for different choices of $\omega$. The oscillations are not present for $\omega < 1$ as the low-frequency signal behaves more like a constant term. The oscillations are most dominant at $\omega \approx 2$, and are quickly suppressed for higher frequencies. The model with $\omega = 10$ shows only percent-level oscillations. High-frequency models cannot be easily measured in data given the suppression of their signal due to the smoothing by the mass kernel. \textit{Bottom:} the same as the top panels but now fixing $\omega = 8$ and varying halo mass instead. The amplitude of the smooth, non-oscillatory part of the signal changes with halo mass, but the amplitude of the oscillations themselves is nearly constant across mass.}
    \label{fig:OscBias_freq_scan}
\end{figure}

We have thus far discussed the results from the QSFosc model with $\omega = 1.5$. Figure \ref{fig:OscBias_combined} now shows the measurements and corresponding theory predictions for the same model but with $\omega \in \{3, 4.5\}$, which are twice and three times that of the fiducial frequency. First, we see that the oscillations are consistently found in the halo bias measurements for a range of frequencies, and the theory model adequately fits them in all cases. As was found in Section \ref{sec:results:bias}, the measurements from the higher mass bins are well fit by the model when using the universality-based predictions for $b_\phi$ and $b_\sigma$. We emphasize that the same coefficients can adequately fit the data across the different models. This is expected because the definition of $b_\phi$ and $b_\sigma$ (which are responses of $n_h$ and $d\ln\nu/dM$ to the large-scale fluctuation) are independent of the bispectrum model.

One noteworthy result from Figure \ref{fig:OscBias_combined} is that the amplitude of the halo-bias oscillations is dependent on the primordial signal's frequency. That is, while our input primordial model has a fixed amplitude regardless of frequency, the late-time oscillations in the halo bias depend strongly on the frequency of this model. This dependence comes about by contrasting the two aspects of the calculation: the wavenumber support of the halo window function and the number of times the squeezed-limit primordial bispectrum oscillates within that window. If there are multiple oscillations, then the signals average down towards zero and the amplitude of halo bias oscillations is significantly suppressed.

While Figure \ref{fig:OscBias_combined} illustrates this behavior using simulated measurements and the associated theory predictions, we also use the theory model to extend this demonstration to a much wider range of frequencies, $\omega$. The top panels of Figure \ref{fig:OscBias_freq_scan} show the expected bias for halos at $M = 10^{14.7} \msun/h$, for different $\omega$ choices. In the discussion to follow, we will refer to specific values of $\omega$ where a given behavior is present or absent in our measurements, but note that these statements also depend on the range of modes, $k$, being utilized.

\begin{figure}
    \centering
    \includegraphics[width=\columnwidth]{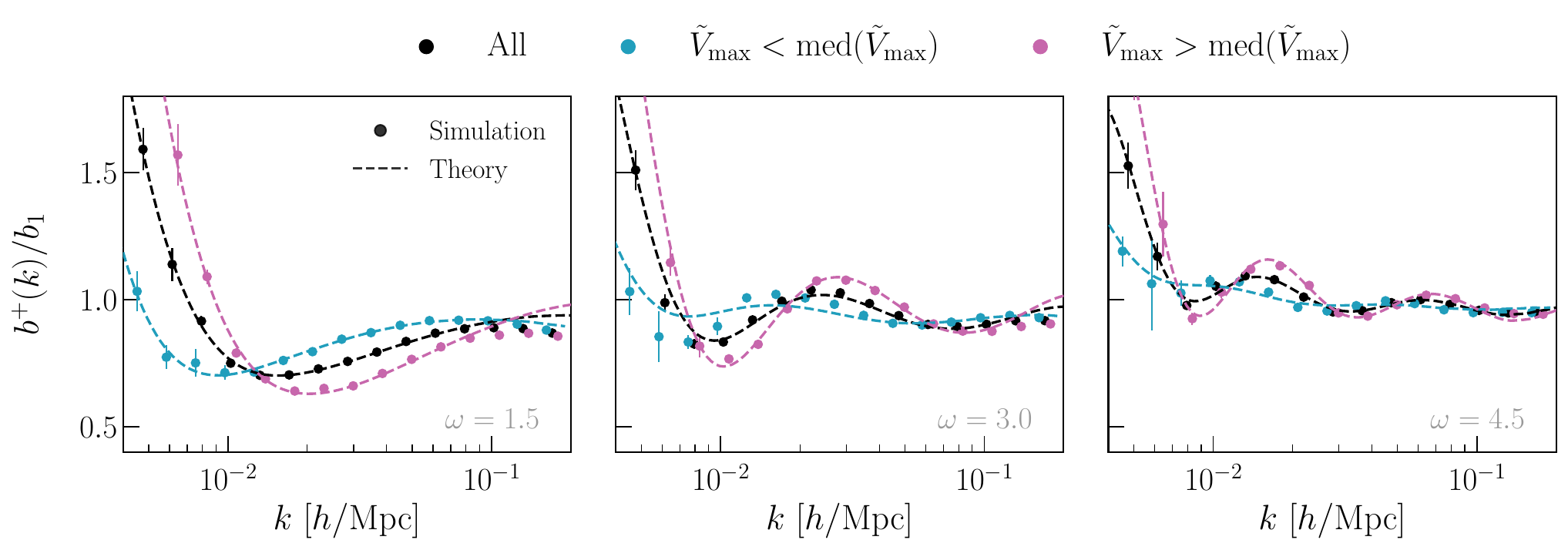}
    \caption{Similar to Figure \ref{fig:OscBias_Vmax} but for higher-frequency models as well. We show results for just $M = 10^{14.7} \msun/h$ for brevity. There is a clear $\tVmax$ dependence in the phases, for all frequencies, that can be fit using our model. The values of $b_\phi$ and $b_\sigma$ needed for the fit are similar across all frequencies (Figure \ref{fig:bphi_bsigma}).}
    \label{fig:OscBias_Vmax_combined}
\end{figure}

For $\omega < 1$, the primordial oscillation is simply a near-constant offset and does not generate a halo-bias oscillation within the presented $k$-range; if we extended the range to $k_{\rm min} = 0.001 \,h/\mpc$, the oscillation emerges.
The oscillation behavior (in our chosen scale range) is dominant at $\omega \approx 2$ and still visible up to $\omega = 4.5$. For $\omega > 8$, the effects are reduced to a few percent across all presented scales. The bottom panels of Figure \ref{fig:OscBias_freq_scan} show the behavior for a fixed model, $\omega = 8$, with varying halo masses. We expect a halo-mass dependence in the amplitude of the signal given it scales with $b_\phi \propto (b_1 - 1)$, which is both mass dependent and will flip from positive to negative at sufficiently low mass. This is precisely the behavior seen in Figure \ref{fig:OscBias_freq_scan}. Note, however, that this predominantly impacts the smooth, non-oscillatory component of the signal.\footnote{While $b_\phi$ scales $\mathcal{F}_R$, which includes both non-oscillatory and oscillatory terms, it does not scale the derivative term in Equation \eqref{eqn:theory:delta_b}. This latter term is a significant contributor to the halo bias signal at $M \lesssim 10^{14} \msun/h$ for $z = 0$, and $M \lesssim 10^{13} \msun/h$ for $z = 1$. Thus, for lower-mass halos, changes to $b_\phi$ are not sufficient for significantly impacting the amplitude of the halo-bias oscillations.} The amplitude of the oscillations, on the other hand, is largely consistent across halo mass.

Finally, Figure \ref{fig:OscBias_Vmax_combined} shows that the assembly bias effect is also observed consistently across different frequencies. The model is able to accurately fit the behavior of these selections. There are somewhat larger residuals for the low-$\tVmax$ bin in the $\omega = 3$ model. However, we can adequately fit all other selections, including the low-$\tVmax$ selection at lower/higher frequencies, so we do not consider the larger residuals as a sign of a problem with the theory model. Following our discussions above, we find the different selections can be fit by the same $b_\phi$ and $b_\sigma$ coefficients across the different $\omega$ choices. The values of these best-fit coefficients are presented in Appendix \ref{appx:bias}. Our results indicate that the currently calibrated $b_\phi$ models/methods, developed for use in Local-type PNG analyses \citep[e.g.][]{Dalal:2026:bphi,Sullivan:2025:bphi,Nascimento:2026:bphi}, can also be utilized in those of collider-type PNGs.

\section{Conclusions}\label{sec:conclusions}

We have produced simulations with a physically motivated cosmological collider signal and present the first measurements of the non-monotonic, oscillating halo bias that arises from such signals. Our simulations use an updated method for generating initial conditions from a given bispectrum, with a particular focus on squeezed-limit accuracy, and this allows us to accurately capture the signals of interest in this work. The simulated halo samples and density fields enable tests of the standard PNG modeling tools in the literature, and their extension from Local-type PNG analyses to those of collider PNGs.

Our main findings are as follows:
\begin{itemize}
    \item The halo bias clearly exhibits oscillations in the linear regime, with the amplitude and phase of the oscillations having a specific mass dependence. The latter is predicted accurately using our theoretical model (Figure \ref{fig:OscBias}). The model has two coefficients, $b_\phi$ and $b_\sigma$, whose values can be theoretically determined for mass-selected halo samples.

    \item The phase of the oscillations also shows a clear dependence on assembly bias --- at fixed mass, more concentrated halos (determined via $\tVmax$) have oscillations shifted to smaller scales than their less-concentrated counterparts (Figure \ref{fig:OscBias_Vmax}). As was the case before, our model can adequately fit these data, with both coefficients now being a function of $\tVmax$.

    \item These results generalize to higher frequencies --- we can measure and fit the oscillations precisely (Figure \ref{fig:OscBias_combined}), including in the case of assembly bias-based effects (Figure \ref{fig:OscBias_Vmax_combined}), for a range of frequencies. Interestingly, for a fixed halo sample definition, there is a common set of $b_\phi$ and $b_\sigma$ values that can fit the measurements from different PNG models (Figure \ref{fig:bphi_bsigma}). This indicates that existing calibrations of $b_\phi$, made to study Local-type PNGs, can be utilized for studies of collider PNGs as well.

    \item Using our theoretical model, we show that bispectra with higher-frequency oscillations result in a weaker oscillation signal in the halo bias, as the Fourier-space oscillations average down within the halo window function (Figure \ref{fig:OscBias_freq_scan}). This suppression is present across all halo masses and demonstrates a limit to the range of frequencies that are observationally viable.
\end{itemize}

Our findings have key implications for future work, particularly for the feasibility of observational constraints on these collider models. The scale-dependent bias of the original Local-type PNG was rapidly identified as having a unique observational signature --- it did not mimic any feature produced by gravitational structure formation, and therefore could be reliably modeled. Instead, a key challenge has been on the measurement front where observed samples of galaxies can exhibit spatial correlations on large scales that are not cosmological in origin and instead arise from survey observing strategies and contamination from the Galactic plane \citep[\eg][]{Monroy2022DESLSS, Kong:2025:DESI, Anbajagane:2025:Balrog, Chaussidon:2025:DESI}. In this work, we identify the oscillations in the halo bias as another unique signal that is not mimicked by gravitational structure formation but is also not easily mimicked by observational systematics. While it is possible for the latter to conspire and induce a non-monotonic feature (i.e., an oscillation) across the scales of interest, it is far more challenging for such systematics to induce multiple oscillations of a coherent frequency, and to also induce the mass-dependent phases/amplitudes present in these signatures. Thus, the collider signal --- or more generally, non-monotonic features in the squeezed-limit bispectra --- can serve as an even cleaner observational target than the original Local-type signal.

It is also straightforward to incorporate such extended PNG models into existing pipelines for scale-dependent halo bias. The halo bias in such PNG models can be numerically predicted as an integral over the bispectrum, and does not require expensive N-body simulations. The key new addition, relative to the Local-type PNG model, is the parameter $b_\sigma$. The value of this parameter depends on the definition of the galaxy sample and therefore cannot be known a priori. However, unlike the Local-type PNG case, where there is a direct degeneracy between $b_\phi$ and $\fNL$, the signal here is sensitive to the combinations $\fNL b_\phi$ and $\fNL b_\sigma$. While the relative importance of each term depends on the signal and on the halo mass being considered, the presence of both terms can allow for some self-calibration opportunities as the scale-dependent behavior of each term varies, and so their impact may be distinguishable. We leave a detailed, survey-focused study of this self-calibration to future work.

Finally, the methods that we present are instrumental for pursuing collider-based studies in ongoing ground-based surveys such as the Dark Energy Spectroscopic Instrument \citep{DESI2016Science} and Rubin Observatory \citep{LSST2018SRD}, and space-based missions such as SPHEREx \citep{Dore:2014:SphereX}, \textit{Roman} \citep{Spergel2013}, and \textit{Euclid} \citep{Euclid}. There is also the potential for millimeter-wave surveys to enhance these constraints further \citep[\eg][]{Hill:2013:tSZPNG, Anbajagane:2026:tSZ}, given their sensitivity to the statistics of massive halos. Observational constraints from any of these data will require reliable modeling prescriptions, as well as simulation suites for testing and mock analyses. Our work presents a key step forward in generating robust, high-precision simulations of collider models --- and primordial bispectra at large --- to enable richer analyses of early Universe physics.

\section*{Acknowledgements}

DA was supported by the National Science Foundation (NSF) Grant No. AST-2508321. DA is also grateful to the Perimeter Institute's visitor program, which led to the conception of this work. Research at Perimeter Institute is supported in part by the
Government of Canada through the Department of Innovation, Science and Economic Development Canada and by
the Province of Ontario through the Ministry of Economic Development, Job Creation and Trade. This work was completed in part with resources provided by the University of Chicago’s Research Computing Center.  

All analysis in this work was enabled greatly by the following software: \textsc{Pandas} \citep{Mckinney2011pandas}, \textsc{NumPy} \citep{vanderWalt2011Numpy}, \textsc{SciPy} \citep{Virtanen2020Scipy}, and \textsc{Matplotlib} \citep{Hunter2007Matplotlib}. We have also used
the Astrophysics Data Service (\href{https://ui.adsabs.harvard.edu/}{ADS}) and \href{https://arxiv.org/}{\texttt{arXiv}} preprint repository extensively during this project and the writing of the paper.

\section*{Data Availability}

Our pipeline for generating initial conditions---including our new technique for approximating PNG kernels using binning---is available at \url{https://github.com/DhayaaAnbajagane/Aarambam}. The simulations are publicly released as part of the \textsc{Ulagam} suite. More details can be found at \url{https://ulagam-simulations.readthedocs.io}. The released products include the 3D density field and halo catalogs for all ten snapshots from each model. Please contact DA if you are interested in other products from the simulations.

\bibliographystyle{mnras}
\bibliography{References}



\appendix

\section{Fits to bias coefficients}\label{appx:bias}

\begin{figure}
    \centering
    \includegraphics[width=\columnwidth]{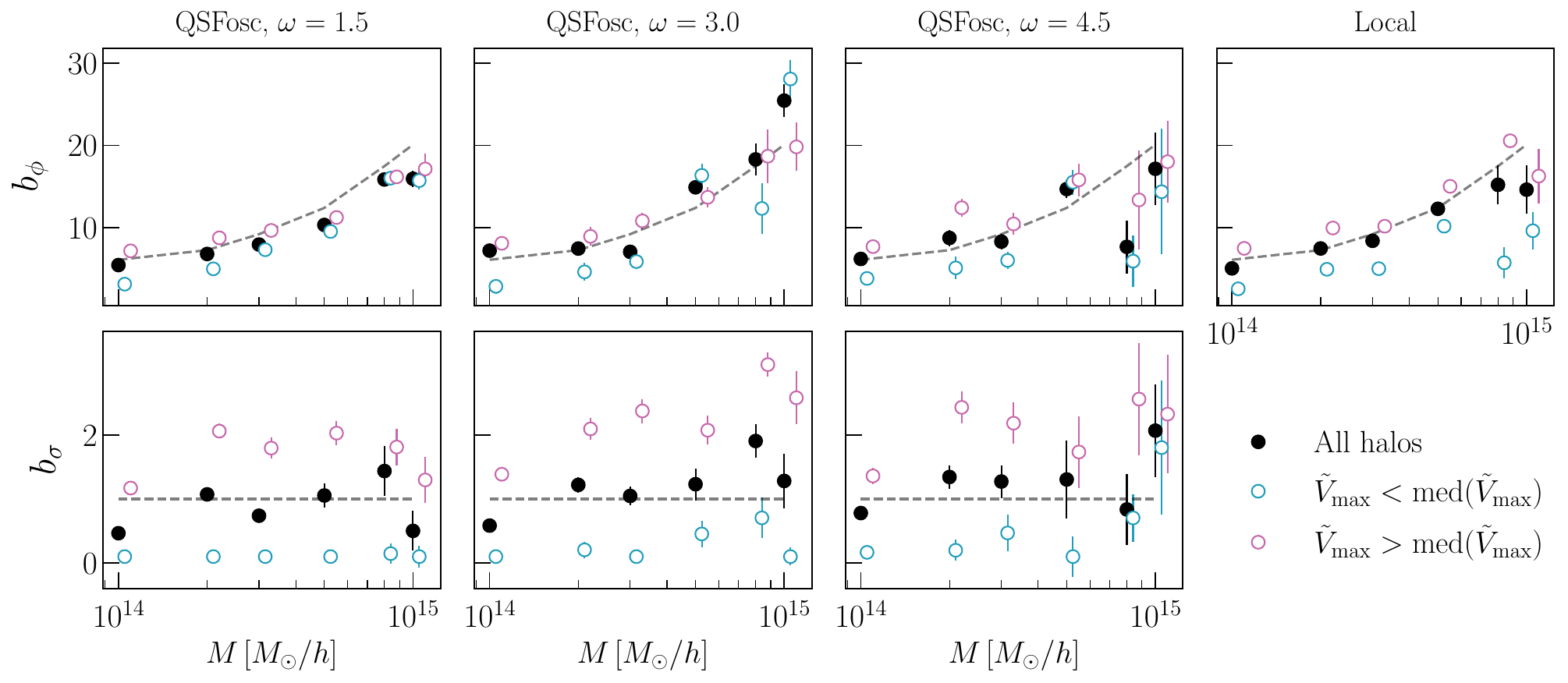}
    \caption{The best-fit bias values, $b_\phi$ and $b_\sigma$, for measurements of the different models (columns) and different $\tVmax$-split samples. The gray lines are the universality prediction for each parameter and closely match the measurements from the mass-selected sample. There are deviations after imposing $\tVmax$ selections but these are largely consistent across all models being considered. The lowest-mass bin deviates slightly from universality even for the mass-selected sample as the halos are resolved by only 50 particles, and therefore have a halo concentration-based selection as well. We do not fit $b_\sigma$ for the Local-type PNG as the associated term in Equation \eqref{eqn:theory:delta_b} is zero by definition.}
    \label{fig:bphi_bsigma}
\end{figure}

Figure \ref{fig:bphi_bsigma} lists our best-fit $b_\phi$ and $b_\sigma$ coefficients as a function of mass bin, PNG model, and $\tVmax$ selection. We overplot gray lines that indicate the universality-based predictions, $b_\phi = \delta_c (b_1 - 1)$ and $b_\sigma = 1$. The mass-based selection is within 10\% to 20\% of the universality prediction, and this is true for all models. The different $\tVmax$ selections result in similar shifts across models for both bias parameters. Thus, interestingly, the halo bias predictions for different models can be well fit using a mostly common set of parameters. This can be expected as these bias parameters represent a property of the halo sample, rather than of the PNG model itself, and so if the halo selection is fixed then the parameter values will be largely similar. This also means methods that have been developed to calibrate $b_\phi$ for Local-type PNGs can be readily adopted for collider PNGs as well.

\section{Validation on Local-type PNGs}\label{appx:local}

\begin{figure}
    \centering
    \includegraphics[width=\columnwidth]{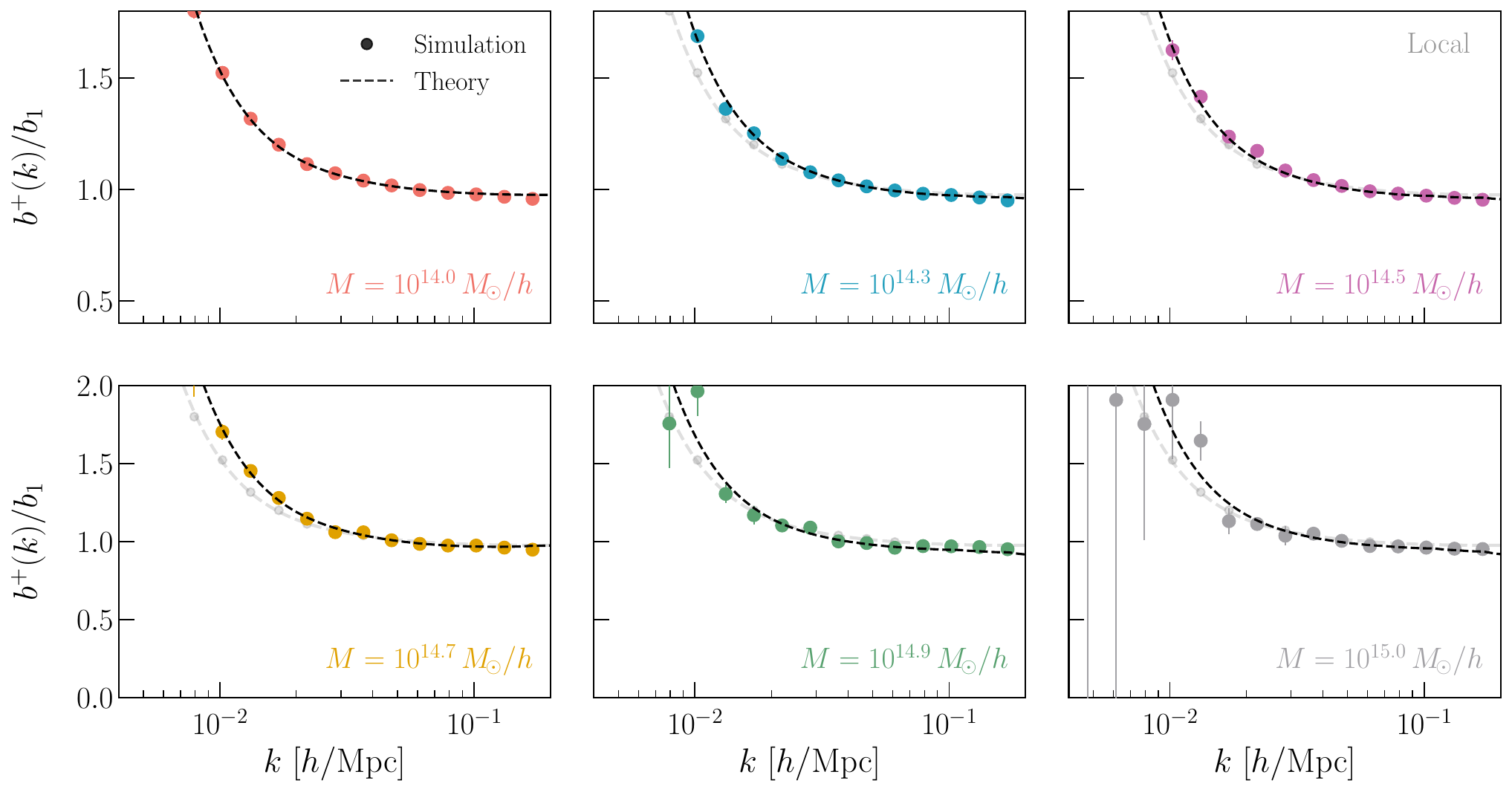}
    \caption{Similar to Figure \ref{fig:OscBias} but for the Local-type PNG model. There is good agreement between theory and measurements for all halo masses.}
    \label{fig:LocalBias}
\end{figure}

As part of our validation, we use the same initial conditions method in Section \ref{sec:sims:ICs} to simulate the Local-type PNG model. Figure \ref{fig:LocalBias} shows the resulting measurements of halo bias for the six mass bins considered in our main analysis, alongside the theory predictions using Equation \eqref{eqn:theory:delta_b}. For the Local-type PNG, the derivative term in Equation \eqref{eqn:theory:delta_b} vanishes as the mass-dependence in $\mathcal{F}_R$ cancels. The precise match between the measurement and the prediction in Figure \ref{fig:LocalBias} is a validation of our methodology. We have also checked that the initial conditions --- for the Local-type PNG model --- from our binned basis function approach are consistent, within numerical accuracy, when compared with those from \citetalias{paper1}, and from the \textsc{2LPTPNG} codebase \citep{Coulton2022QuijotePNG}.

\newpage


\label{lastpage}
\end{document}